\documentclass[10pt, conference]{IEEEtran}

\usepackage{amsthm}

\usepackage{subfig}
\usepackage{tikz}
\usetikzlibrary{matrix}

\newcommand\supp{\mathop{\rm supp}}

\usepackage[utf8x]{inputenc}
\usepackage[T1]{fontenc}
\usepackage{lmodern}

\newcommand{\p}{.\,}
\newcommand{\G}{\ensuremath{\mathcal{G}}}
\newcommand{\D}{\ensuremath{\mathcal{D}}}

\newcommand{\exc}{\textup{exc}}

\newcommand{\B}{\ensuremath{\mathcal{B}}}

\newcommand{\HH}{\ensuremath{\mathcal{H}}}

\usepackage{enumerate}

\usepackage{amsmath}
\usepackage{amssymb}

\usepackage{graphicx}
\usepackage{url}

\IfFileExists{cmll.sty}
	{\usepackage{cmll}
	 }
	{\newcommand*\coh{\sim}
	 
	 \newcommand*\incoh{{\not\coh}}}

\newcommand*\ii{\ensuremath{\varphi}}

\renewcommand{\le}{\ensuremath{\leqslant}}
\renewcommand{\ge}{\ensuremath{\geqslant}}

\newcommand*\Nn{\ensuremath{\mathbb{N}}}

\newcommand*\Ee{\ensuremath{\mathbb{E}}}

\newcommand*\boldd{\ensuremath{\boldsymbol{d}}}

\newtheorem{define}{Definition}
\newtheorem{cor}{Corollary}
\newtheorem{thm}{Theorem}
\newtheorem{lem}{Lemma}
\newtheorem{prop}{Proposition}

\begin{document}

\title{Convergence law for hyper-graphs\\ with prescribed degree sequences}

\author{
    \IEEEauthorblockN{Nans Lefebvre}
    \IEEEauthorblockA{
        LIAFA, Universit\'e Paris VII\\
        \url{nans.lefebvre@liafa.univ-paris-diderot.fr}
        }
}

\maketitle

\begin{abstract}
We view hyper-graphs as incidence graphs,  i.e. bipartite graphs with a set of nodes representing vertices and a set of nodes representing hyper-edges, with two nodes being adjacent if the corresponding vertex belongs to the corresponding hyper-edge.
It defines a random hyper-multigraph specified by two distributions, one for the degrees of the vertices, and one for the sizes of the hyper-edges.
We develop the logical analysis of this framework and first prove a convergence law for first-order logic, then characterise the limit first-order theories defined by a wide class of degree distributions.
Convergence laws of other models follow, and in particular for the classical Erd\H{o}s-R\'enyi graphs and $k$-uniform hyper-graphs.
\end{abstract}

\section{Introduction}
Understanding the graphs that occur in nature is an important challenge in computer science, as many phenomena can be seen as diffusion processes on graphs, such as the diffusion of ideas in society, from marketing to social uproar, or the spreading of epidemics \cite{britton}. 
Most natural graphs have a sparse structure yet are locally highly clustered, a property that classical random graphs fail to model.
This motivated the introduction of a myriad of new models with more realistic features, with an overview of rigorous results in \cite{durrett}.

This paper considers the bipartite configuration model \cite{blanchet},
interpreted as the incidence graph of a hyper-multigraph by considering one set of vertices to be hyper-edges, and the binary relation to specify that a vertex belongs to a hyper-edge.
A random hyper-graph is specified by two given degree distributions, one for the degrees of the vertices and one for the sizes of the hyper-edges (communities), 
and generates the uniform incidence graph on these degree distributions. 
The analogy with the real-world formation of graphs is compelling, since vertices are often linked by some shared community, actors may be linked because they play in the same film, scientists because they collaborate on a paper \cite{latapy,liu}.
In society, communities can be of any size, involving just two vertices (friendships), a small number of vertices (families), up to very large communities (professions, urban areas).
Hence this is a flexible model that generates a wide class of graphs by allowing a wide range of limit distributions, generates graphs with a sparse structure and dense communities, and models complex relations (see \cite{ghoshal} for an application of hyper-graph as abstractions of social networks).

The logical analysis is one way to understand a random structure, by studying the probabilities that first-order formulas are satisfied. 
It has been started on random graphs independently in \cite{gleb} and \cite{fagin76}, a result celebrated as the zero-one law of random graphs. 
The whole of the Erd\H{o}s-R\'enyi model was successfully charted \cite{spencer0strange}, and \cite{telles} extended results to \emph{$k$-uniform} hyper-graphs with a different logical formalism, using a $k$-ary predicate.
The logical structure traditionally used allows only to treat simple graphs; following \cite{courcelle}, it is shown that using the incidence graph as logical structure allows to handle non-uniform hyper-multigraphs without modifying the logic, using just a binary predicate.
This allows to study directly the logic of random hyper-multigraphs, and since this framework extends conservatively the classical one it allows to recover previous results on classical Erd\H{o}s-R\'enyi graphs and $k$-uniform hyper-graphs as special cases.

The main result is a convergence law for a class of random non-uniform hyper-multigraphs (Theorem \ref{thm_conv}). 
Besides natural conditions to avoid pathological sequences, the sequences in this class must have limit distributions with a bounded second moment and a sublinear fourth moment.
This includes the case of Dirac distributions that generate hyper-graphs with uniform hyper-edge sizes.
It extends a result of Lynch \cite{lynch05} that gave a convergence law for a class of graphs defined by their degree sequence, and contributes to the body of works extending classical results to hyper-graphs, such as \cite{elie} for the phase transition.
Two sequences of random hyper-multigraphs are contiguous if they share the same limit theory, i.e. first-order formulas are either almost surely true on both sequences or neither, allowing to compare random sequences and models from a logical point of view.
The possible limit theories are characterised and axiomatised, and from this the contiguity of random hyper-multigraphs is deduced (Theorem \ref{thm_cont}).

In the second section, the logical framework and notations are introduced. In the third section, the considered class of random hyper-multigraphs is defined, and the convergence law is proved for this class. 
In the fourth section, the limit theories are investigated and a criterion for the contiguity of random sequences is deduced; then some distributions are compared to other models, such as Erd\H{o}s-R\'enyi graphs.
In the fifth section, some properties of interest are studied, and some remarks on future work conclude.

\section{Preliminaries}
In this section, the most general objects are defined first and more classical structures such as simple graphs are defined as specific cases.
This is consistent with the logical model since graphs can be defined by conditioning on the validity of a first-order property.
Some definitions may be non-standard to fit the context of this paper.

Let $(n)_k$ denote the $k$-falling factorial of $n$, $(n)_k = \prod_{i=0}^{k-1}(n-i)$.
If $\mathcal{S}$ is a (countable) set, let $\wp(\mathcal{S})$ denote the set of probability distributions on $\mathcal{S}$, $\boldsymbol{S}$ be an element of $\wp(\mathcal{S})$, and $s \sim \boldsymbol{S}$ be a random variable sampled from $\boldsymbol{S}$.

\subsection{Hyper-multigraphs and the incidence graph}
A (non-uniform, undirected) \emph{hyper-multigraph} $H = (V = \{1, \dots, n\}, M: \mathcal{P}(V) \mapsto \Nn)$ is a set of vertices $V$ equipped with a set of hyper-edges given with multiplicities.
For a set $e \in \mathcal{P}(V)$, $e \in E$ denotes that $M(e) > 0$, and we suppose that any $e$ with $M(e)>0$ is of size at least $2$.
A hyper-multigraph  is a \emph{hyper-graph} if all multiplicities are $1$, in which case the hyper-graph can be considered as a structure $H = (\{1, \dots, n\}, E)$.
A hyper-multigraph is \emph{$k$-uniform} if every hyper-edge is of size $k$, and is a \emph{multigraph} if it is $2$-uniform.
A \emph{graph} is a $2$-uniform hyper-graph.
Let the classes of all hyper-multigraphs, hyper-graphs, multigraphs, and graphs be denoted respectively by $\mathbb{H}_\Nn$, $\mathbb{H}$, $\mathbb{G}_\Nn$, and $\mathbb{G}$.
\emph{Random hyper-multigraphs} (elements of $\wp(\mathbb{H}_\Nn)$) are denoted by $\HH$, and random hyper-multigraph sequences by $(\HH_n)_{n \in \Nn}$.

Two vertices $v$ and $v'$ are \emph{adjacent}, denoted by $v \coh v'$, if there is an $e \in E$ such that $v \in E$ and $v' \in E$.
The converse notion, that there is no $e \in E$ such that $v \in e$ and $v' \in e$, is denoted by $v \incoh v'$.
The degree of a vertex $v$, denoted $\deg(v)$ is the number of edges containing $v$, i.e. $|\{e \mid e \in E \wedge v \in e\}|$.
These two notions can be specialised to $(m,k)$-adjacency by specifying respectively the number $m$ of hyper-edges, and the size $k$ of the edges,
i.e. $v \coh^m_k v'$ if there are $m$ hyper-edges of size $k$ containing $v$ and $v'$.

The following definitions can be extended to hyper-multigraphs, but will only be used on simple graphs, usually denoted by $G$. 
The \emph{excess} $\exc(G)$ of a graph $G$ is $|E|- |V|$.
A \emph{homomorphism} of a graph $g=(V_g,E)$ to $G$ is an injection $\sigma: V_g \mapsto V$ such that for all $(v_1, v_2) \in V_g^2$, $v_1 \coh v_2$ implies $\sigma(v_1) \coh \sigma(v_2)$. 
A graph $H$ is a \emph{subgraph} of $G$ if there is a homomorphism from $V_H$ to a subset of $V_G$, and this subset is called a \emph{realisation} of $H$.
A subgraph is (asymptotically) \emph{realisable} in a random graph $\G$ if it is (asymptotically) realised probability bounded away from $0$.
A homomorphism is an \emph{isomorphism} if it is bijective and $(v_1, v_2) \in V_g^2$, $v_1 \coh v_2 \Leftrightarrow\sigma(v_1) \coh \sigma(v_2)$.
A \emph{rooted graph} is a graph with a distinguished vertex (the root), and a rooted graph isomorphism is a graph isomorphism that maps the roots together.
A graph $G$ induces a distance between vertices defined as $d_G(x,y) = \min \{t \mid \exists x_1, \dots,  x_{t+1} \textup{ such that } x_1 \coh \dots \coh x_{t+1} \wedge x_1 = x \wedge x_{t+1} = y\}$ (by convention if this number is not defined it is considered to be infinite).
A \emph{ball} $B(r, l) = \langle V_B, E_B, r\rangle$ of \emph{root} $r$ and \emph{radius} $l$ is the subgraph of $G$ defined by $V_B = \{ v \in V_G \mid d(v, r) \le l \}$ and $E_B$ is the restriction of $E_G$ to pairs of vertices in $V_B$.
A \emph{neighbourhood} of a graph denotes a ball of finite radius, and its \emph{type} denotes its isomorphism class. 

A \emph{bipartite graph} is a graph with two sets of vertices $V$ and $W$, such that for all $(v,v') \in V^2$ (respectively  $W^2$), $v \incoh v'$. 
Classes of bipartite hypergraphs, etc., can be defined, but in this paper only bipartite simple graphs are used as in the next definition:
\begin{define}
The incidence graph $I_G$ of a hyper-multigraph $G = (V,M)$ is a bipartite graph $(V,E_I, \in_I)$ with $E_I = \{(e,i) \mid i \le M(e) \}$ such that $v \in e \leftrightarrow v \in_I (e,i)$. Each hyper-edge is represented by a node in $E_I$, and $\in$-edges  link vertices to the hyper-edges they belong to.
\end{define}
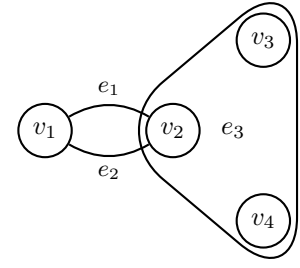
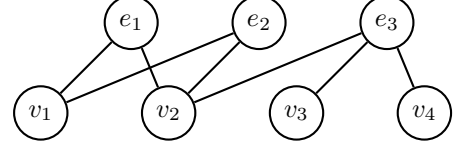
\begin{figure}%
    \centering
    \subfloat[A small hyper-multigraph with a multiple edge and a hyper-edge]{
        \begin{tikzpicture}[-,auto,node distance=1.7cm,thick,main node/.style={circle,draw}]
          \node[main node] (1) {$v_1$};
          \node[main node] (2) [right of=1] {$v_2$};
          \node[main node] (3) [above right of=2] {$v_3$};
          \node[main node] (4) [below right of=2] {$v_4$};   

          \path[every node/.style={font=\sffamily\small}]
            (1)         edge [bend right] node[below] {$e_2$} (2)
            (2)         edge [bend right] node[above] {$e_1$} (1);
          
          \draw [rounded corners=10mm] (0.8,0)--(3.35,-2.15)--(3.35,2.15)--cycle;
          \node at (2.5, 0) {$e_3$};
         \end{tikzpicture}}%
    \qquad
    \subfloat[$\in$-edges are implicitly oriented]{
        \begin{tikzpicture}[-,auto,node distance=1.7cm,thick,main node/.style={circle,draw}]
          \node[main node] (1) {$e_1$};
          \node[main node] (2) [right of=1] {$e_2$};
          \node[main node] (3) [right of=2] {$e_3$};
          \node[main node] (4) [below left of=1] {$v_1$};
          \node[main node] (5) [right of=4] {$v_2$};
          \node[main node] (6) [right of=5] {$v_3$};
          \node[main node] (7) [right of=6] {$v_4$};   

          \path[every node/.style={font=\sffamily\small}]
            (1)         edge node {} (4)
            (1)         edge node {} (5)
            (2)         edge node {} (4)
            (2)         edge node {} (5)  
            (3)         edge node {} (5)
            (3)         edge node {} (6)   
            (3)         edge node {} (7);      
        \end{tikzpicture}}%
    \caption{A hyper-multigraph and its incidence graph}%
    \label{fig:ex1}%
\end{figure}
\begin{define}
Let $\mathbb{I}$ denote the class of incidence graphs, that is the set of bipartite graphs with an ordered pair of sets of vertices where every vertex in $E_I$ has degree at least $2$.
\end{define}
The incidence graph of two hyper-multigraphs are isomorphic if and only if the two hyper-multigraphs are isomorphic, therefore:
\begin{prop}
The class of hyper-multigraphs $\mathbb{H}_\Nn$ and the class of incidence graphs $\mathbb{I}$ are in bijection.
\end{prop}
This last property allows to use the incidence graph as a faithful representation of hyper-multigraphs.

\subsection{The bipartite configuration model}
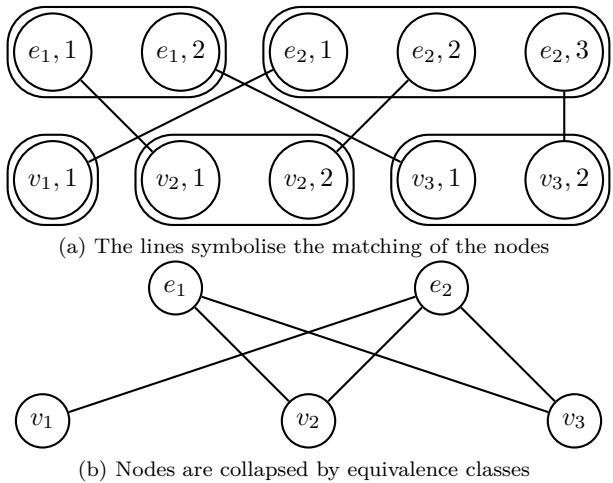
\begin{figure}%
    \centering
    \subfloat[The lines symbolise the matching of the nodes]{
        \begin{tikzpicture}[-,auto,node distance=1.7cm,thick,main node/.style={circle,draw}]
          \node[main node] (1) {$e_1, 1$};
          \node[main node] (2) [right of=1] {$e_1, 2$};
          \node[main node] (3) [right of=2] {$e_2, 1$};
          \node[main node] (4) [right of=3] {$e_2, 2$};
          \node[main node] (5) [right of=4] {$e_2, 3$};
          \node[main node] (6) [below of=1] {$v_1, 1$};
          \node[main node] (7) [right of=6] {$v_2, 1$};
          \node[main node] (8) [right of=7] {$v_2, 2$};
          \node[main node] (9) [right of=8] {$v_3, 1$};   
          \node[main node] (10)[right of=9] {$v_3, 2$};

          \path[every node/.style={font=\sffamily\small}]
            (1)         edge node {} (7)
            (2)         edge node {} (9)
            (3)         edge node {} (6)
            (4)         edge node {} (8)  
            (5)         edge node {} (10);     
         \draw [rounded corners=5mm]  (-0.6,-0.6)--(2.3,-0.6)--(2.3,0.6)--(-0.6,0.6)--cycle;
         \draw [rounded corners=5mm]  (2.8,-0.6)--(7.4,-0.6)--(7.4,0.6)--(2.8,0.6)--cycle;
         \draw [rounded corners=5mm]  (-0.6,-1.1)--(0.6,-1.1)--(0.6,-2.3)--(-0.6,-2.3)--cycle;
         \draw [rounded corners=5mm]  (1.1,-1.1)--(4,-1.1)--(4,-2.3)--(1.1,-2.3)--cycle;
         \draw [rounded corners=5mm]  (7.4,-1.1)--(4.5,-1.1)--(4.5,-2.3)--(7.4,-2.3)--cycle;
        \end{tikzpicture}}%
    \qquad
    \subfloat[Nodes are collapsed by equivalence classes]{
        \begin{tikzpicture}[-,auto,node distance=2.5cm,thick,main node/.style={circle,draw}]
          \node[main node] (1) {$v_1$};
          \node[main node] (2) [above right of=1] {$e_1$};
          \node[main node] (3) [below right of=2] {$v_2$};
          \node[main node] (4) [above right of=3] {$e_2$};
          \node[main node] (5) [below right of=4] {$v_3$}; 
          \path[every node/.style={font=\sffamily\small}]
            (1)         edge node {} (4)
            (3)         edge node {} (4)
            (5)         edge node {} (4)
            (2)         edge node {} (3)  
            (2)         edge node {} (5);
        \end{tikzpicture}}%
    \caption{A configuration and corresponding incidence graph}%
    \label{fig:ex2}%
\end{figure}

The bipartite configuration model $\B(d^v, d^e)$ is a way to generate random hyper-multigraphs by specifying their corresponding incidence graphs.
To generate an incidence graph on $n$ vertices, this model takes two degree sequences, $d^v=(d^v_0, d^v_1, \dots)$ and $d^e = (d^e_0, d^e_1, \dots)$,
such that $d^v(k)$ (respectively $d^e(k)$) is the number of vertices in $V$ (respectively in $E$) of degree $k$, with $S = \sum d^v = \sum d^e$.
A configuration $C = \langle f^v: C_V \mapsto V, f^e: C_E \mapsto E, \sigma \rangle$ is then given by two sets of nodes $C_V = C_E = \{1, \dots, S\}$, two partitions according to $d^v$, $d^e$, and a random permutation $\sigma$ of $\{1, \dots, S\}$
(equivalently, a random matching of $C_V$ to $C_E$).
The incidence graph is then obtained by collapsing nodes in $C_V$ into a vertex of degree $k$ according to $f^v$ and collapsing nodes of $C_E$ according to $f^e$,
and setting $v \in e$ if and only if there exist $c_v$, $c_e$ such that $f(c_v) = v$, $f(c_e) = e$, and $\sigma(c_v) = c_e$.
This process may not define an incidence graph since there may be multiple edges between pairs of vertices in $V$ and $E$ - in this case, it does not define a proper hyper-multigraph since hyper-edges are sets of vertices.
However, with some conditions on the sequence of degree sequences (that the sequences considered here satisfy), it generates an incidence graph with positive probability \cite{blanchet}, and the random structure obtained by conditioning on the absence of multiple $\in$-edges is the uniform hyper-multigraph with degree sequences $(d^v, d^e)$.
Figure \ref{fig:ex2} show a configuration and the corresponding graph with $d^e(2)=1$, $d^e(3)=1$, $d^v(1)=1$, $d^v(2)=2$ and all other degrees and edge sizes set to $0$.
Let $\D$ denote a sequence of degree sequences, i.e. $\D = (d_1, d_2, \dots)$ where each $d_n$ is a sequence of natural numbers (therefore $d_n(k)$ is the number of vertices of degree $k$ in the $n$-th graph).
Let $\D'^v$ (resp. $\D'^e$) be the sequence $(p^v_1, p^v_2, \dots)$ with $p^v_n(k) = \frac{1}{|V|} d^v_n(k)$ (resp. $p^e_n(k) = \frac{1}{|E|} d^e_n(k)$).
Then for each pair $(\D^v= (d^v_1, d^v_2, \dots), \D^e= (d^e_1, d^e_2, \dots))$ of sequences, $\B(\D^v, \D^e)$ denotes the random incidence graph sequence on the corresponding degree sequences.
By definition, incidence graphs have no edge of size smaller than $2$, therefore it is assumed that $d^e_n(0)=d^e_n(1) = 0$ for all $n \in \Nn$.
We consider pairs of sequences of sequences that converge weakly to a pair of \emph{asymptotic degree distributions}, denoted $(\boldd^v,\boldd^e)$, i.e.  $\D^v \leadsto \boldd^v$ and $\D^e \leadsto \boldd^e$.
See \cite{billingsleyconv} for details on the weak convergence of distributions, it implies that for all $k$ 
$$\lim_{n \rightarrow \infty} p^v_n(k) = \boldd^v(k).$$
Note that there may be no graph that realises a given degree sequence; however with some conditions on the limit distributions (that distributions considered here satisfy), asymptotically there is always a graph realising the degree sequence \cite{cbcoursrg}.
Another coherence condition is assumed, that the sums $\sum_k k d^v_n(k)$ and $\sum_k k d^e_n(k)$ must be equal; however the number of edges is adjusted depending on the number of vertices, so the average of $\boldd^v$ may be different from the average of $\boldd^e$.

\subsection{Logic}
The language of first-order (FO) logic on graphs is defined as the closure under boolean connectives and quantifiers of equality and a binary predicate.
Usually the variables are interpreted as vertices and the binary predicate is interpreted as the adjacency relation $\coh$, however doing so makes inexpressible any property about edge multiplicities or hyper-edges.
Instead, having as set of variables the union of vertices and edges and interpreting the binary predicate as the $\in$ relation on the incidence graph defines FO$_\in$.
Formally, the logical structure is $\langle V \cup E_I, \in \rangle$.
This allows to handle hyper-multigraphs without modifying the logical language.
\begin{prop}
Any graph property is FO-definable if and only if it is FO$_\in$-definable.
\end{prop}
See \cite{courcelle}. Therefore all results on the FO-logic of graphs are true on the FO$_\in$-logic of graphs (consequently, FO$_\in$ is denoted FO).

A \emph{sentence} is a formula without free variables, and the \emph{quantifier-depth} of a formula is defined as the maximal nestedness of a quantifier in the formula.
A structure $H$ satisfies a formula $\ii$, denoted $H \models \ii$, if $\ii$ is true on $H$.
For a sentence $\ii$ and a sequence of random structures $(\HH_n)_{n \in \Nn}$, let $\mu(\ii, \HH_n)$ denote $\Pr[H_n \models \ii]$ with $H_n \sim \HH_n$, and let $\mu(\ii, \HH)$ denote the limit of this sequence (called \emph{limiting probability}) if it exists.
A sequence of random structures $(\HH_n)$ has a \emph{convergence law} (for FO-logic) if every sentence $\ii$ has a limiting probability $\mu(\ii, \HH)$ (it is a zero-one law if all limiting probabilities are either $0$ or $1$).
A sentence is (asymptotically) almost surely true if its limiting probability is $1$, and an \emph{axiom} is such a sentence.
The \emph{theory} of a set of axioms is the set of all sentences implied by the axioms.
\begin{define}
The \emph{limit theory} (or \emph{almost sure} theory) $T_\HH$ of a random structure $\HH$ is the set of its almost surely true sentences. A \emph{limit hyper-multigraph} is a countable hyper-multigraph that is a \emph{model} of the theory, i.e. it satisfies all formulas of the limit theory.
\end{define}
A limit theory is consistent and so always has a model, and unless the theory is complete (which happens only if there is a zero-one law)  
there are infinitely many non-isomorphic models. 
These definitions can naturally be specialised to graphs, and bipartite graphs.

To see the expressive power of FO on hyper-multigraphs, some syntactic sugar can be defined. 
For example, even if there is only one domain, vertices can be distinguished from edges by a formula, i.e. $x$ is a vertex if and only if $v(x)$ is true, with $v(x) \equiv \forall y \p \neg (y \in x)$.
The formula characterising edges is naturally the negation of $v(x)$, i.e. $e(x) \equiv \exists y \p (y \in x)$. This allows to define edge and vertex quantifiers as $\exists^V x \equiv \exists x \p v(x)$ and $\exists^E x \equiv \exists x \p e(x)$.
That $u$ and $v$ are adjacent can be expressed by $u \coh v \equiv \exists e \p (u \in e \wedge v \in e)$, and the size of an edge can be expressed by 
$|e|=2 \equiv \exists v_1 v_2 \forall u \p (v_1 \in e \wedge v_2 \in e \wedge v_1 \neq v_2 \wedge ((u \neq v_1 \wedge u \neq v_2) \implies u \notin e))$
where $v \notin e$ denotes $\neg( v \in e)$.
Similarly, note that $|e|=k$, as well as any $\coh_k(v_1, \dots, v_k)$ or $(m,k)$-adjacency relation, is definable by a first-order formula.
Furthermore, it is possible to define subclasses of hyper-multigraphs in first-order logic:
\begin{prop}
The subclasses $\mathbb{G}_\Nn$, $\mathbb{G}$ and $\mathbb{H}$ of $\mathbb{H}_\Nn$ are FO-definable.
\end{prop}
A hyper-multigraph is a multigraph if every edge is of size $2$, and a graph if there are no multiple edges. This can be expressed by a first-order formula
$\Phi_\mathbb{G} = (\forall^E e \p (|e|=2)) \wedge \neg(\exists e_1 e_2 \exists u v \p (u \in e_1 \wedge v \in e_1 \wedge u \in e_2 \wedge v \in e_2 \wedge e_1 \neq e_2)) $
while $\Phi_{\mathbb{G}_\Nn} = (\forall^E e \p (|e|=2))$.
Therefore, for all $H\in \mathbb{H}_\Nn$, $H \models \Phi_\mathbb{G}$ if and only if $H \in \mathbb{G}$.
A hyper-multigraph is a hyper-graph (is in $\mathbb{H}$) if there are no multiple hyper-edges, which can be expressed by the formula
$\neg(\exists e_1 \exists e_2 \forall v \p (e_1 \neq e_2 \wedge (v \in e_1 \leftrightarrow v \in e_2))$.
It follows that convergence laws can be deduced on subclasses by interdefinability:
\begin{prop}
Let $\mathcal{H}$ be a random hyper-multigraph having a convergence law. Then the random graph $\G$ defined as the distribution induced by $\mathcal{H}$ on graphs has a convergence law, with 
$$\mu(\ii, \G) = \frac{\mu(\ii\wedge \Phi_\mathbb{G} , \HH)}{\mu(\Phi_\mathbb{G}, \HH)}.$$
\end{prop}

\section{The random hyper-multigraph with given degree sequences}
In this section, we define some restrictions on the class of sequences we consider, which allow us to prove that this class satisfies properties that imply a convergence law.
After the definitions and statements, the proof is given first assuming these properties.
Then in the following subsection the theorem is proven for trivial pairs of distributions, to simplify subsequent arguments. 
The following subsections prove that the properties hold with high probability, 
using a key lemma showing that the expected number of realisations of small subgraphs can be computed from the configurations. 

To get a convergence law, it is necessary to put some restrictions on the sequences considered.
Three conditions encompassing a large class of degree sequences are given, and entail three properties of the corresponding sequences of graphs. 
The main theorem is then that the random sequence defined by the three former conditions correspond exactly to the theories satisfying the three latter properties.

The considered sequences of random incidence graphs satisfy, :
\begin{enumerate}[(i)]
\item $\D'^v$ and $\D'^e$ converges weakly to $\boldd^v$ and $\boldd^e$ (in $\wp(\Nn)$).
\item There exists a $N$ such that $\forall n>N$, $\forall k \in \Nn$, $\boldd^v(k)=0$ implies that $d^v_n(k) = 0$ and $\boldd^e(k)=0$ implies that $d^e_n(k) = 0$.
\item For $(\D, \boldd)$ being $(\D^v,\boldd^v)$ or $(\D^e,\boldd^e)$, 
\begin{equation}
\lim_{n \rightarrow \infty}  \sum_{i = 1}^{n} \frac{d_n(k)}{n} k^2 = \sum k^2 \boldd(k) < \infty \tag{$*$}\label{eq_h2}
\end{equation}
\begin{equation}
 \sum_{i = 1}^{n} \frac{d_n(k)}{n} k^4 = o(n). \tag{$**$}\label{eq_h4}
\end{equation}
\end{enumerate}
Condition (i) signifies that the degree sequences converge to probability distributions on $\Nn$. 
Condition (ii) signifies there is a threshold after which all represented degrees and edge sizes have to be in the support of the distributions.
Condition (iii) signifies that the distributions have a bounded second moment (\ref{eq_h2}) and that their tails are not too heavy (\ref{eq_h4}), i.e. their fourth moment grows slower than $n$. 
Note that there are no requirements on the pair of distributions, so the distribution $\boldd^e$ is completely independent of $\boldd^v$. 

Examples of distributions satisfying these properties are distributions with finite support (the case of Dirac distributions generating hyper-multigraphs with uniform edge sizes or regular degrees)
or power law distributions, i.e. $d(k) \approx c \cdot k^{-\alpha}$ with $\alpha$ greater than $3$. The power laws with exponent smaller than $3$ do not have a sublinear fourth moment so do not satisfy condition (iii). 
A sequence that does not satisfy (ii) may have $0$ vertices of degree $k$ for even $n$ and $1$ vertex of degree $k$ for odd $n$, violating the possibility of a convergence law since having degree $k$ is a first-order property (note that such a sequence can still satisfy condition (i) and (iii), and therefore still have a local weak limit).

The considered graph sequences satisfy:
\begin{enumerate}[(I)]
\item Every acyclic neighbourhood has either no realisations or an unbounded number of realisations.
\item The number of realisations of every unicyclic neighbourhood converges to a finite value.
\item Every subgraph with positive excess has asymptotically no realisation.
\end{enumerate}
Note that these three properties cover all types of subgraphs, since a connected graph has excess $-1$ if and only if it is a tree, excess $0$ if it is unicyclic, and positive excess if it has more than one cycle.  

These properties shed more light on conditions (i)-(iii).
The conditions (i)-(iii) have two natures, the first is to avoid pathological sequences (condition (ii)), and the second one is to ensure that properties (I)-(III) hold (condition (iii)).
It may be proven that there is a convergence law even when relaxing conditions of the second sort, but in that case the limit theory will be qualitatively different.
Suppose that for all $n>N$, the sequence has $m$ vertices of degree $k$ for some $N,m \in \Nn$ (therefore $k \notin \supp(\boldd^v)$). 
Then condition (I) is violated, as there are exactly $m$ neighbourhoods with a vertex of degree $k$ (and with high probability these neighbourhoods are all trees).
For condition (iii), note that when the tails of the distributions get heavier, it means that there are more and more vertices of high degrees, and at some threshold these vertices account for such an important proportion of edges that they get connected to each others, realising subgraphs of positive excess, violating (III).

\begin{lem}\label{lem_conv}
Let $\B(\D^v,\D^e)$ be an incidence graph sequence satisfying (i)-(iii). Then it satisfies (I)-(III).
\end{lem}
This allows to state the main theorem:
\begin{thm}\label{thm_conv}
Let $\B(\D^v,\D^e)$ be an incidence graph sequence satisfying (i)-(iii). It defines a sequence of random hyper-multigraphs that has a convergence law.
\end{thm}
In the following subsection, the proof is given assuming that Lemma \ref{lem_conv} holds, as the truth of first-order formulas is determined from neighbourhoods alone. 
Then the theorem is proven for the special cases of trivial pairs of distributions, allowing to simplify subsequent arguments.
The next subsections prove that properties (I)-(III) hold, thereby proving Lemma \ref{lem_conv}.

\subsection{Proof of the convergence law}
\tikzset{
  arn_n/.style = { circle, draw=black,
    fill=black},
}
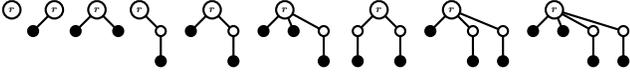
\begin{figure}
\begin{center}
\begin{tikzpicture}[-,auto,thick,main node/.style={circle,draw},scale=0.4, every node/.style={scale=0.4}]]
\node[main node] (1) {$\boldsymbol{r}$};
\node[arn_n] (3) [below right of=1] {};
\node[main node] (2) [above right of=3] {$\boldsymbol{r}$};
\node[arn_n] (4) [below right of=2] {};
\node[main node] (5) [above right of=4] {$\boldsymbol{r}$};
\node[arn_n] (6) [below right of=5] {};

\node[main node] (1') [above right of=6] {$\boldsymbol{r}$};
\node[main node] (11') [below right of=1'] {};
\node[arn_n] (12') [below  of=11'] {};
\node[arn_n] (3') [ right of=11'] {};
\node[main node] (2') [above right of=3'] {$\boldsymbol{r}$};
\node[main node] (21') [below right of=2'] {};
\node[arn_n] (22') [below of=21'] {};
\node[arn_n] (4') [right of=21'] {};
\node[main node] (5') [above right of=4'] {$\boldsymbol{r}$};
\node[arn_n] (6') [right of=4'] {};
\node[main node] (7') [right of=6'] {};
\node[arn_n] (8') [below of=7'] {};

\path[every node/.style={font=\sffamily\small}]
(2) edge node {} (3)
(4)	edge node {} (5)
(6)	edge node {} (5)

(1') edge node {} (11')
(11') edge node {} (12')
(2') edge node {} (3')
(2') edge node {} (21')
(21') edge node {} (22')
(4')	edge node {} (5')
(6')	edge node {} (5')
(7')	edge node {} (5')
(7')	edge node {} (8')
;
\end{tikzpicture}
\begin{tikzpicture}[-,auto,thick,main node/.style={circle,draw},scale=0.4, every node/.style={scale=0.4}]]
\node[main node] (1'')  {$\boldsymbol{r}$};
\node[main node] (11'') [below right of=1''] {};
\node[arn_n] (12'') [below of=11''] {};
\node[main node] (13'') [below left  of=1''] {};
\node[arn_n] (14'') [below of=13''] {};
\node[arn_n] (3'') [right of=11''] {};
\node[main node] (2'') [above right of=3''] {$\boldsymbol{r}$};
\node[main node] (21'') [below right of=2''] {};
\node[arn_n] (22'') [below  of=21''] {};
\node[main node] (23'') [right of=21''] {};
\node[arn_n] (24'') [below of=23''] {};
\node[arn_n] (4'') [right of=23''] {};
\node[main node] (5'') [above right of=4''] {$\boldsymbol{r}$};
\node[arn_n] (6'') [right of=4''] {};
\node[main node] (7'') [right of=6''] {};
\node[main node] (8'') [right of=7''] {};
\node[arn_n] (9'') [below of=7''] {};
\node[arn_n] (10'') [below of=8''] {};

\path[every node/.style={font=\sffamily\small}]
(1'') edge node {} (11'')
(11'') edge node {} (12'')
(1'') edge node {} (13'')
(13'') edge node {} (14'')
(2'') edge node {} (3'')
(2'') edge node {} (21'')
(2'') edge node {} (23'')
(21'') edge node {} (22'')
(23'') edge node {} (24'')
(4'')	edge node {} (5'')
(6'')	edge node {} (5'')
(7'')	edge node {} (5'')
(8'')	edge node {} (5'')
(7'')	edge node {} (9'')
(8'')	edge node {} (10'')
;
\end{tikzpicture}
\end{center}
\caption{Types of $3$-spheres}
\label{fig:spheres}
\end{figure}
To prove the convergence law, first the notion of $q$-sphere is defined inductively. 
It is a finite subclass of the class of balls of radius $q-1$ introduced to capture the expressive power of first-order logic. 
There is only one $1$-sphere, where the root has no neighbours. Then a $(q+1)$-sphere is a root with at most $q$ neighbours of each type of $q$-sphere. 
By definition there is only a finite number of types of $q$-spheres.
See Figure \ref{fig:spheres}, where black vertices mean that the vertex has no other neighbour.
Intuitively, each quantifier allows to go further, and 
The neighbourhood of a vertex $v$ can be mapped to a $q$-sphere by setting (inductively) to $q$ the number of neighbours of $v$ that are of some type of $(q-1)$-sphere if there are more than $q$ such neighbours.
Hanf locality lemma (see the standard textbook \cite{elementsfmt}) states that two structures $G_1$, $G_2$ are equivalent up to $q$-quantifiers sentences if for every type $T$ of $(3^q)$-sphere, both $G_1$ and $G_2$ have more than $q$ neighbourhoods of type $T$ or both have the same number of neighbourhoods of type $T$.
By properties (I)-(III), for any $q$ the number of neighbourhoods of each type is either unbounded, or converges to a finite value. It follows that every first-order sentence has a limiting probability.

\subsection{Trivial pairs of distributions}
To simplify subsequent arguments, in the other subsections it is assumed that $\supp(\boldd^v)$ is different from $\{0\}$, $\{1\}$, and $\{0,1\}$, i.e a pair of distribution is \emph{trivial} if  $\supp(\boldd^v) \subseteq \{0,1\}$. 
Otherwise it would be necessary to add particular cases to every statement, so these pairs are treated as special cases here.
Note that if this is the case, there is a convergence law (even a zero-one law), albeit a fairly uninteresting one; 
in particular, the number of unicyclic neighbourhoods is zero for every such neighbourhood, while for non-trivial distributions it converges to a non-trivial value. 
If $\supp(\boldd^v) = \{0\}$, then the set of almost sure formulas are simply formulas stating that there exist vertices, and formulas about non-adjacency. 
The first set gives an axiom scheme, and the second set follows from the axiom $ \neg (\exists e \exists v \p v \in e)$. 
There is, up to isomorphism, only one countable graph that models these axioms (the countable empty graph).
Now if $1 \in \supp(\boldd^v)$, there are hyper-edges in the graph, but they are all isolated, i.e. $\neg (\exists v \exists e_1 e_2 \p v \in e_1 \wedge v \in e_2 \wedge e_1 \neq e_2)$. 
There is an axiom scheme for every edge size $k$ in the support of $\boldd^e$, stating that there are at least $m$ such hyper-edges.
There is, up to isomorphism, only one countable graph that models these axioms, a countable collection of isolated hyper-edges of size $k$ for each $k \in \supp(\boldd^e)$, along with a countable collection of isolated vertices if $0 \in \supp(\boldd^v)$.
Note that these arguments hold for every possible $\boldd^e$.
From model-theoretical arguments, the uniqueness of the countable model of the limit theory implies a zero-one law, and therefore convergence.

\subsection{Proof of property (I): the local weak limit}
The property (I) is proved using the local weak limit; the idea is to look at the `typical' neighbourhood of a vertex. With high probability this neighbourhood is a tree, and the tree observed by picking a vertex uniformly at random converges to a simple random object, a branching process. This branching process gives that any tree that can be realised is realised infinitely often. 

A graph $G$ induces a probability distribution $\mu_G$ on rooted graphs by taking a uniform root among all vertices, i.e.
$$\mu_G = \frac{1}{|V|} \sum_{v \in V}\delta(G,v),$$ 
where $\delta$ is the Dirac distribution over the isomorphism class of the rooted graph $(G, v)$.
A sequence of random graphs $(\G_n)$ has weak local limit $\mu$ if $\mu_{\G_n} \leadsto \mu$, i.e. the sequence $\mu_{\G_n}$ converges weakly to $\mu$ (for details on weak convergence in metric spaces, see \cite{billingsleyconv}).
\begin{figure}
\begin{center}
	\begin{tikzpicture}
		\matrix (a) [matrix of math nodes, column sep=0.2cm, row sep=0.2cm]{
		\textup{Root} 			& & & \bullet & & V \\
		\textup{Generation 1} 	& & \bullet & & \bullet & E \\
		\textup{Generation 2} 	& \bullet & \bullet & \bullet & \bullet & V \\
								& & & \dots & &  \\
		};
		\foreach \i/\j in { 1-4/2-3, 1-4/2-5, 2-3/3-2, 2-3/3-3, 2-3/3-4, 2-5/3-5} 
		\draw (a-\i) -- (a-\j);
	\end{tikzpicture}
\end{center}
\caption{First generations of a bipartite branching process}
\label{fig:branch}
\end{figure}
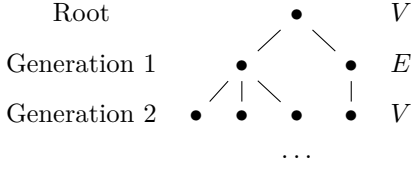
A branching process is a random rooted tree, defined by a root (generation $0$), with each vertex in generation $i$ having a random number of offspring vertices in generation $i+1$.
Let $BGW(\boldd^v, \boldd^e)$ be a two-sorted Galton-Watson branching process, where each point has an independent number of offsprings. It is defined inductively by a root of type $V$ (respectively $E$), with $x \sim \boldd^v$ offsprings of type $E$ (respectively $x \sim \boldd^e$ offsprings of type $V$). Then a point in generation $i>0$ has $x \sim \boldd^*$ offsprings, with $\boldd$ being $\boldd^v$ if the point is of type $V$ and $\boldd^e$ if it is of type $E$, with
$$\boldd^*(x)  = \frac{(x+1)\boldd(x+1)}{\sum_{y}y \boldd(y)}$$
(this accounts for the edge already present between the point and its parent). An example is given in Figure \ref{fig:branch}.

\begin{lem}\label{lem_lwc}
If $\B(\D^v,\D^e)$ satisfies (i)-(iii), then $\mu_\B \leadsto BGW(\boldd^v, \boldd^e)$.
\end{lem}
The proof is based on the exploration of the connected component containing the root, and can be adapted from the standard proof of convergence for the configuration model (see \cite{mezmon} for details). 
It allows to deduce property (I) from the following consequence:
\begin{cor}
Let $\B(\D^v,\D^e)$ satisfy (i)-(iii) and $T$ be a bipartite tree, the following are equivalent: 
\begin{itemize}
\item $T$ is asymptotically realised infinitely often.
\item $T$ is realisable.
\item $\forall v \in V_T$ (resp. $\forall e \in E_T$), $\deg(v) \in \supp(\boldd^v)$ (resp. $\deg(e) \in \supp(\boldd^e)$).
\end{itemize}
\end{cor}
This corollary follows directly from the previous lemma, since the probability that $\B(\D^v,\D^e)$ is isomorphic to a tree $T$ is given by $BGW(T)$. 
If this probability is positive, then there is a linear number of vertices in $\B$ with neighbourhood isomorphic to $T$. 
Since $BGW(T)>0$ if only if all of the degrees appearing in $T$ are in the support of the distributions $(\boldd^v,\boldd^e)$, property (I) follows.

\subsection{Subgraph counting lemma}
This subsection gives a formula for the expected number of small subgraphs and proves it. 
It proves the convergence of the expected number of realisations of subgraphs with non-negative excess, proving the remaining two properties in the next subsection.

Recall that $(n)_k$ denotes the $k$-falling factorial of $n$.
The expected number of realisations of a bipartite graph with non-negative excess can be computed from the following formula:
\begin{lem}\label{lem_con}
Let $H$ be a (connected) subgraph with $k_v$ vertices, $k_e$ edges, $m$ $\in$-edges, $c_v$ its number of vertex-automorphisms, $c_e$ its number of edge-automorphisms.
Let $d^v \sim \boldd^v$, $d^e \sim \boldd^e$ be random variables, and recall that $\exc(H) = m -(k_v + k_e)$. 
Then the expected number of realisations of $H$ is 
$$n^{-\exc(H)} \frac{\prod \Ee[(d^v)_{\deg_H(v_i)}]  \cdot \prod \Ee[(d^e)_{\deg_H(e_j)}]}{c_vc_e (\Ee d^v)^{m-k_e}(\Ee d^e)^{k_e}}.$$
\end{lem}
Here a vertex (resp. edge) automorphism is an automorphism that maps vertices to vertices (resp. edges to edges).
Note that this is the expectations of the number of embeddings, so if $H$ has degrees that are not in the support of the distribution but are smaller than degrees in the support, this number will be positive. 
Contrarily, if $H$ contains some degree $l$ bigger than any degree in the support, then the expectation of the $l$-falling factorial is $0$ hence the whole expression goes to $0$.

This formula is proven directly from the configuration.  
The expected number of realisations of $H$ can be written as $\frac{1}{c_vc_e}\sum_{\tau}\Ee[\rho(\tau(H),G)]$, 
where the sum is over all injective maps from $V_H$ to $V_G$ and $E_H$ to $E_G$, and $\rho$ is the number of homomorphisms for a given map, i.e. $\rho$ is positive if $\tau$ is a homomorphism and $0$ otherwise.
Recalling that $S$ is the number of $\in$-edges, there are $(n)_{k_v} \cdot (\frac{S}{\Ee d^e})_{k_e}$ such maps, as there are $n = \frac{S}{\Ee d^v}$ vertex equivalence classes, and $\frac{S}{\Ee d^e}$ edge equivalence classes.
The division by the number of automorphisms accounts for the repetitions of maps. 
The first $\in$-edge of the homomorphism between vertex $v_1$ and edge $e_1$ is matched with probability $\deg(\tau(v_1)) \cdot \deg(\tau(e_1))/S$, 
as there are $\deg(\tau(v_1))$, $\deg(\tau(e_1))$ nodes in the configuration that can be chosen to create the $\in$-edge between $v_1$ and $e_1$, and it is chosen among $S$ nodes. 
For the following edges, the number of nodes corresponding to vertices that have already been matched is correspondingly diminished, and similarly the number of nodes becomes $(S-i)$ if $i$ $\in$-edges have already been matched. 
Define the product
$$P_V = \prod_{i = 1}^{k_v} (\deg_G(\tau(v_i)))_{\deg_H(v_i)},$$
and similarly $P_E$ where the sum is over all edges.
Now, a uniformly sampled set of $k$ vertices has the same distribution as $\boldd^{\bigotimes k}$, the $k$-th product of $\boldd$ 
(to see this, the $k$ vertices are distinct with probability $(n)_k/n^k$ which goes to $1$ as $n$ goes to infinity).
Therefore 
$$\Ee \prod_{i=1}^{k} (d_i)_{k_i} =  \prod_{i=1}^{k} \Ee[(d)_{k_i}]$$
allowing to rewrite $P_V$ as $P'_V = \prod_{i = 1}^{k_v} (d^v)_{\deg_H(v_i)}$ (and similarly $P_E$ as $P'_E$).
Now we have $\lim_{n \rightarrow \infty} (S-k) /n = \Ee d^v$ for any $k \in \Nn$, and therefore $(S)_k \approx n^k(\Ee d^v)^k$.
Putting all this together, we have 
$$\frac{1}{c_vc_e (S)_m} \sum_\tau (P_V \cdot P_E) = \frac{n^{k_v + k_e}(\Ee d^v)^{k_e}}{c_vc_e (S)_m (\Ee d^e)^{k_e}} (P'_V \cdot P'_E).$$
By rewriting $(S)_m$ by $n^m(\Ee d^v)^m$, we obtain a factor $n^{k_v + k_e - m}$, i.e. $n^{-\exc(H)}$.
Simplifying the expression by this rewriting yields the expected formula.

\subsection{Proof of properties (II) and (III): subgraphs of non-negative excess}
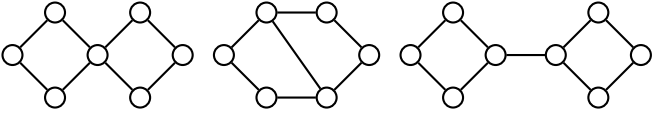
\begin{figure}%
    \centering
        \subfloat{
        \begin{tikzpicture}[-,auto,thick,main node/.style={circle,draw, scale=0.8}]
          \node[main node] (1){};
          \node[main node] (2) [above right of=1]{};
          \node[main node] (3) [below right of=1]{};
          \node[main node] (4) [below right of=2]{};
          \node[main node] (5) [above right of=4]{};
          \node[main node] (6) [below right of=4]{};          
          \node[main node] (7) [below right of=5]{};

          \path[every node/.style={font=\sffamily\small}]
            (1)         edge node {} (2)
            (1)         edge node {} (3)
            (2)         edge node {} (4)
            (3)         edge node {} (4)
            (4)         edge node {} (5)
            (4)         edge node {} (6)
            (5)         edge node {} (7)
            (6)         edge node {} (7);      
        \end{tikzpicture}%
        }
       \hspace{.01cm} 
      \subfloat{
      \begin{tikzpicture}[-,auto,thick,main node/.style={circle,draw, scale=0.8}]
      \node[main node] (1){};
      \node[main node] (2) [above right of=1]{};
      \node[main node] (3) [below right of=1]{};
      \node[main node] (4) [right of=2]{};
      \node[main node] (5) [right of=3]{};
      \node[main node] (6) [below right of=4]{};

      \path[every node/.style={font=\sffamily\small}]
        (1)         edge node {} (2)
        (1)         edge node {} (3)
        (2)         edge node {} (4)
        (2)         edge node {} (5)
        (3)         edge node {} (5)
        (4)         edge node {} (6)
        (5)         edge node {} (6);
        \end{tikzpicture}%
        }
	\hspace{.01cm}
        \subfloat{
        \begin{tikzpicture}[-,auto,thick,main node/.style={circle,draw, scale=0.8}]%
          \node[main node] (1){};
          \node[main node] (2) [above right of=1]{};
          \node[main node] (3) [below right of=1]{};
          \node[main node] (4) [below right of=2]{};
          \node[main node] (4')[right of=4]{};
          \node[main node] (5) [above right of=4']{};
          \node[main node] (6) [below right of=4']{};          
          \node[main node] (7) [below right of=5]{};

          \path[every node/.style={font=\sffamily\small}]
            (1)         edge node {} (2)
            (1)         edge node {} (3)
            (2)         edge node {} (4)
            (3)         edge node {} (4)
            (4)         edge node {} (4')
            (4')        edge node {} (5)
            (4')        edge node {} (6)
            (5)         edge node {} (7)
            (6)         edge node {} (7);      
        \end{tikzpicture}%
        }
    \caption{Bipartite graphs of excess $1$}%
    \label{fig:exc}%
\end{figure}

Now we prove properties (II) and (III).
To prove property (III), note that any graph with positive excess contains a subgraph of excess $1$ which can be of three different forms, either it is constituted of two cycles sharing a vertex and its maximal degree is $4$, or it is constituted of two cycles linked by a path, or a cycle with a chord, and in both the maximal degree is $3$ (see Figure \ref{fig:exc}).
To deduce property (III), it is sufficient to see that graphs with excess $1$ are not realisable:
\begin{cor}
Let $\B(\D^v,\D^e)$ be a random incidence graph sequence satisfying (i)-(iii) and $H$ be a bipartite graph with $\exc(H)=1$. Then $H$ is not realisable. 
\end{cor}
First note that by condition (iii), the average degree converges to a finite value, so it suffices to show that for a graph with positive excess, the value given by Lemma \ref{lem_con} goes to zero. 
Since this value is multiplied by $n^{-\exc(H)} = n^{-1}$, it is sufficient to show that the numerator is in $o(n)$, which is exactly the second part of condition (iii) since the maximal degree of a subgraph of excess $1$ is either $3$ or $4$.
Then any graph with positive excess contains an embedded subgraph of excess $1$ which is not realisable, so it is not realisable itself and property (III) follows. 

When applied to unicyclic graphs (graphs with excess $0$), the formula given by Lemma \ref{lem_con} gives property (II): 
\begin{cor}
Let $\B(\D^v,\D^e)$ be a random incidence graph sequence satisfying (i)-(iii) and $H$ be a bipartite graph with $\exc(H)=0$, then the expected number of realisations of $H$ converges. 
\end{cor}
By the first part of condition (iii), the numerator is finite, therefore every term in the expression is finite. 
There is in fact convergence in distribution to Poisson random variables of the number of cycles for each size of cycles (there is even joint convergence). 
Details can be found in \cite{cbcoursrg,jansonmulti2}.
This concludes the proof of convergence.

\section{Limit theories and first-order contiguity}
In this section, we investigate the limit theories defined by the incidence graphs sequences $\B(\D^v,\D^e)$.
First, an axiomatisation of the limit theories is given, and the models of the limit theories are investigated.
Then, as some sequences have the same limit theories, we look at sequences through the lens of first-order contiguity, a tool to compare random sequences from the point of view of logic.
Then classical models of random graphs are analysed as special cases of this framework.
The completions of the limit theories are studied next, and in the last subsection the models of the limit theories are investigated. 

\subsection{First-order axioms}
An axiomatisation of the limit theory can be deduced from properties (I)-(III) and distributions $(\boldd^v, \boldd^e)$.
Let $T_{\boldd^v, \boldd^e}$ be the limit theory of $\B(\D^v, \D^e)$.
First, by property (II) this axiomatisation contains no sentence about unicyclic neighbourhoods, since each may be realised or not. 
Then property (III) gives that every sentence stating the absence of neighbourhoods of positive excess is in the limit theory.
For all $(r,s) \in \Nn^2$, let $Z_{r,s}$ be the formula stating `for every set of $r$ vertices and $s$ edges, there are not $r+s+1$  $\in$-edges', and $Z$ be the axiom scheme of all such sentences.
Then, for all $k \notin \supp(\boldd^v)$ (resp. $k \notin \supp(\boldd^e)$), let $Y^v_k$ (resp. $Y^e_k$) be the formula stating `there is no vertex of degree exactly $k$', and $Y$ denote the axiom scheme of all $Y^v$ and $Y^e$ sentences. 
Note that this axiom scheme can be replaced by a single sentence if both supports of the distributions are finite or cofinite. 
Then, for all bipartite trees $T$, and all $m \in \Nn$, let $X^T_m$ be the formula stating `there are at least $m$ exact realisations of $T$', and $T$ denote the axiom scheme of all such sentences. 
Here, an exact realisation means that for all internal nodes in $T$, the formula states that the vertex has exactly the degree it has in $T$. 
For leaves, the formula states that the vertex has degree $1$ if and only if the node has no offspring in the branching process (since edges must be of size at least $2$, leaves can only be vertices). 
Therefore the formula asserts the existence of $T$ as induced components if and only if the branching process can die out isomorphic to $T$.
The three axiom schemes $X$, $Y$, $Z$ (with $X$ and $Z$ depending on the distributions) axiomatise the almost sure theory of $\B(\D^v, \D^e)$, 
i.e. $T_{X,Y,Z} = T_{\boldd^v, \boldd^e}$.

\subsection{First-order contiguity}
To compare two sequences of random structures from a logical point of view, the notion of first-order contiguity is used, as an analogue to the probabilistic notion of contiguity of \cite{jansonconti}.
\begin{define}
Two sequences of random structures are first-order contiguous if and only if every sentence has a limiting probability either on the two sequences or on neither of them, and both sequences share the same almost sure theory. 
\end{define}
In other words, two sequences of random hyper-multigraphs are first-order contiguous if and only if for every first-order sentence $\ii$, $\ii$ is either not almost sure (in general it may not converge), or it is almost sure on the two sequences.
\begin{thm}\label{thm_cont}
Two sequences of hyper-multigraphs defined respectively by $\B(\D^v_1,\D^e_1)$ and $\B(\D^v_2,\D^e_2)$ both satisfying (i)-(iii)
are first-order contiguous if and only if $(\supp(\boldd^v_1), \supp(\boldd^e_1)) = (\supp(\boldd^v_2), \supp(\boldd^e_2))$.
\end{thm}
Which means that from the point of view of first-order logic, the statistical information given by the distributions does not matter, only the supports matter.
Since there is a convergence law, the first property holds obviously, and the second property is immediate since the axiomatisation only depends on $(\supp(\boldd^v), \supp(\boldd^e))$.
Note that the limiting probability of sentences that are not in the almost sure theory may differ.
In fact, this is the case exactly if the limit distributions differ:
\begin{cor}
There is a formula $\ii$ with $\mu(\ii,\B(\D^v_1,\D^e_1)) \neq \mu(\ii,\B(\D^v_2,\D^e_2))$ if and only if $(\boldd^v_1,\boldd^e_1) \neq (\boldd^v_2,\boldd^e_2)$.
\end{cor}
Since all sentence probabilities are only dependent on the limit distributions, the `only if' part is trivial. 
To show the other direction, suppose the two pairs of distributions have the same pair of supports (otherwise the proof is trivial), and suppose $\boldd^v_1$ is different from $\boldd^v_2$ on values $i$ and $j$ (the case $\boldd^e$ being symmetric).
A formula that asserts the existence of a cycle has a non-trivial probability; then specifying the branching of vertices in the cycle to have exactly $i$ neighbours gives a formula satisfying the requirement.

As a final remark, the set of limit theories defined by the class of degree distributions considered here can be ordered by the supports of the corresponding distributions. 
Indeed, the set of realised subgraphs of a sequence $\B_1(\D^v_1,\D^e_1))$ is included in the set of subgraphs realised by a sequence $\B_2(\D^v_2,\D^e_2))$ if the pair of supports of $\B_1$ is included in the pair of supports of $\B_2$, 
i.e. $(\supp(\boldd^v_1),\supp(\boldd^e_1)) \subseteq (\supp(\boldd^v_2),\supp(\boldd^e_2))$. 
In this sense, there is a maximal theory for this class corresponding to the pair of supports $(\Nn, \Nn)$.

\subsection{The case of the (multi)graph with given degree sequence and the Erd\H{o}s-R\'enyi graphs}
Much work has been done on the configuration model, which generates the uniform simple graph on a given degree sequence $\D$. 
This is a specific case that can be generated by the bipartite graph $\B(\D, \D_{\{2\}})$ where $\D_{\{2\}}$ contains only edges, converging to $\delta_{\{2\}}$ the Dirac distribution with mass on $2$.
Note that in that case, the local weak convergence is a simple Galton-Watson, and that in the Lemma \ref{lem_con} the terms depending on the edges cancel out.
Precisely, $\Ee\boldd^e = \Ee[(\boldd^e)_2] = 2$, so for any realisable subgraph, 
$$\frac{\prod_{j=1}^{k_e} \Ee[(d^e)_{\deg_H(e_j)}]}{(\Ee d^e)^{k_e}} =1.$$
In particular the expected number of multiple edges is therefore $\lambda = (\Ee[(\boldd^v)_2])^2/(4 (\Ee\boldd^v)^2)$, and the uniform graph with degree sequence $\boldd$ is obtained by conditioning on the absence of double edges. Since the number of multiple edges converges in distribution to a Poisson variable, the probability that the graph is simple is the probability of $0$ in $\textup{Poi}(\lambda)$, the Poisson distribution of parameter $\lambda$.

In the Erd\H{o}s-R\'enyi graph $\G(n, c/n)$, each edge is independently present with probability $c/n$.
The degree of a vertex is therefore given by a binomial of parameter $n-1$ and $c/n$, i.e. $\Pr[\deg(v) = k] = {n \choose k}(\frac{c}{n})^{k}(1-\frac{c}{n})^{n-k}$.
So the average degree is $c$, and the degree distribution converges to the Poisson distribution of parameter $c$.
Therefore in the limit there are infinitely many vertices of each degree. 
Then Theorem \ref{thm_cont} implies directly:
\begin{cor}\label{cor_er}
Let $\B(\D, \D_{\{2\}})$ be a sequence of incidence graphs satisfying (i)-(iii) with $\supp(\boldd) = \Nn$. The sequence of graphs it defines, conditioned on being simple, is contiguous to $\G(n, c/n)$.
\end{cor} 
There are no multiple edges in $\G(n, c/n)$ by definition, so it is necessary to condition on simplicity.

\subsection{Completions of the limit theories and limiting probabilities}
The set of possible limit theories share a strong resemblance to the theory of the Erd\H{o}s-R\'enyi graph $\G(n, c/n)$, and they can be completed in a similar fashion.
A completion $\tau$ is a set of sentences such that $T+\tau$ is a complete theory, i.e. every formula (or its negation) belongs to the theory.
A convergence law is equivalent to the fact that there is a set $C$ of completions of the limit theory, such that any two elements of $C$ are mutually inconsistent.
A $q$-completion is a sentence that completes the theory only up to $q$ quantifiers. 
In the present case no completion is finite, however some finite formulas are $q$-completions:
\begin{cor}
Any sentence fixing the number of neighbourhoods in each realisable unicyclic type of $(3^q-1)/2$-sphere is a $q$-completion of the theory of $\B(\D^v, \D^e)$.
\end{cor}
This observation was made in \cite{lynch92} for Erd\H{o}s-R\'enyi graphs and is given here without a proof, as it can be adapted from the original paper without much modification. 
The argument follows from properties (I)-(III), the only difference is that the realisable types depend on the supports of the distributions.

Other results of the paper can similarly be extended to the bipartite configuration model. 
In particular, since formulas with non-trivial limiting values are formulas about unicyclic neighbourhoods, it can be shown that the limiting values of formulas can be written using basic arithmetic operations with the values in $\textup{Poi}(\lambda)$ and $BGW(\boldd^v, \boldd^e)$, for any $\lambda$ given by the formula of Lemma \ref{lem_con}.
Conversely, any real number that can be expressed in such a form is the limit of a first-order formula. 
The automata-theoretic arguments used to show this also imply that the asymptotic probability of a formula (or a formula with a given asymptotic probability) can be computed in polynomial space and that this bound is optimal. Details can be found in \cite{lynch92}.

\subsection{Models of limit theories}
In this subsection, the countable models of the limit theory are investigated. 

By the given axiomatisation, models of the limit theories are constituted of tree components and independent unicyclic components.
However, if there are no vertices of degree less than $2$, then the $BGW$ branching process never dies, and there are no small components. Since there is a giant component and that with high probability this component is unique \cite{molloyreedsize}, the graph is connected with high probability. However, the countable graphs that model the limit theory are \emph{not} connected. 
As observed in \cite{spencer0strange}, this is a most roundabout way to show that connectivity is not a first-order property. 

Consider for example the limit theory of $\B(\delta_2,\delta_2)$, the $2$-regular multigraph. The multigraphs it generates are a collection of cycles. 
However, except for a finite number of vertices, all vertices belong to cycles of size larger than $k$, for any $k$. As the length of the cycles grows, formulas of first-order logic cannot distinguish long cycles from an infinite alternating path of vertices and edges. 
Since they realise the same $q$-spheres, two countable graphs that are collections of infinite paths satisfy the same set of first-order formulas, regardless of the number of infinite paths in each.
Therefore the limit models of $\B(\delta_2,\delta_2)$ are constituted of a collection of infinite paths and a collection of small cycles.

Now consider the limit theory of $\B(\delta_3,\delta_2)$, the $3$-regular multigraph. The multigraphs it generates are connected with high probability.
However as $n$ grows the small cycles grow further apart from each others (since two cycles at fixed distance realises a subgraph of positive excess). 
It follows that in the limit the cycles are in isolated components, and each vertex in the cycles is the root of two infinite trees with $3$-regular vertices and $2$-regular edges.
Therefore the limit models of $\B(\delta_3,\delta_2)$ are constituted of a collection of infinite trees  with $3$-regular vertices and $2$-regular edges and a collection of unicyclic components.

Last consider the limit theory of $\B(\boldd^v, \boldd^e)$ with $1 \in \supp(\boldd^v)$. It generates hyper-multigraphs with induced components of finite size.
By the local weak limit, every realised induced component is realised linearly often, and therefore appears infinitely often in the limit.
Then in the limit there may be any number of infinite components (including none) since, from a first-order point of view, infinite components can be simulated by arbitrarily large finite components. 
Therefore we have:
\begin{prop}
Let $\B(\boldd^v, \boldd^e)$ be a sequence satisfying (i)-(iii). 
If $0,1 \notin \supp(\boldd^v)$ then all components of the limit incidence graph are infinite. 
Otherwise there may be any number of infinite components in the limit.
\end{prop}
The idea is the following. 
To prove the first statement, suppose there is a finite component in the limit, when $0$ and $1 \notin \supp(\boldd^v)$.
Let this finite component be called $H$ and have radius $l$, then for any $q$ greater than $l$ the type of the $q$-sphere corresponding to $H$ is not generated by $BGW(\boldd^v, \boldd^e)$, contradicting previous observations. 
For the second statement, it is sufficient to show that the same types  of $q$-spheres are realised independently of the presence of infinite components.
Let $L_1$, $L_2$ be two countable models of the limit theory of $\B(\boldd^v, \boldd^e)$. 
Suppose that $H$ is an infinite tree component in $L_2$ and that $L_1$ does not contain infinite components. 
For any $q$, let $T$ be a type of $q$-sphere realised in $H$. By hypothesis, $L_2$ satisfy property (I) and therefore $T$ is realised infinitely often in $L_1$, in components of radius greater than $q$. 

\section{Other properties of the model}
In this section, two properties of interest are studied, first the probability that a hyper-multigraph is a hyper-graph, and second that, unlike classical models of random graphs (including the graphs with specified degree sequences), the $\coh$-graph can be a sparse graph with non-trivial clustering coefficient.
\subsection{Probability to be a hyper-graph}
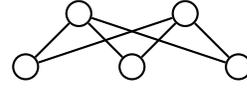
\begin{figure}%
    \centering
        \begin{tikzpicture}[-,auto,thick,main node/.style={circle,draw}]
          \node[main node] (1) {};
          \node[main node] (2) [above right of=1] {};
          \node[main node] (3) [below right of=2] {};
          \node[main node] (4) [above right of=3] {};
          \node[main node] (5) [below right of=4] {}; 
          \path[every node/.style={font=\sffamily\small}]
            (1)         edge node {} (4)
            (3)         edge node {} (4)
            (5)         edge node {} (4)
            (2)         edge node {} (1)
            (2)         edge node {} (3)  
            (2)         edge node {} (5);
        \end{tikzpicture}
    \caption{A triple edge or a double $3$-hyper-edge}%
    \label{fig:multi}%
\end{figure}

Let $k \ge 3$. Then the incidence graph of a double $k$-hyper-edge is a subgraph with $k$ vertices, $2$ hyper-edges, and $2k$ $\in$-edges. 
Therefore the excess of this subgraph is $2k-(2+k)>0$ (see Figure \ref{fig:multi}).
Similarly, the incidence graph of a $k$-multiple edge is a subgraph with $2$ vertices, $k$ edges, and $2k$ $\in$-edges, with excess $2k-(2+k)>0$.
Since subgraphs with positive excess are not realisable, a hyper-multigraph is a hyper-graph if and only if there are no double edges.
Therefore the property to have no multiple hyper-edge can be computed, as it is just the probability to have no $2$-cycle.

Using Lemma \ref{lem_con} to get
$$\lambda = \frac{(\Ee[(\boldd^v)_2])^2 \cdot (\Ee[(\boldd^e)_2])^2}{4 (\Ee\boldd^v)^2(\Ee\boldd^e)^2},$$
the probability of this property is then the probability of $0$ in $\textup{Poi}(\lambda)$.

\subsection{Clustering coefficient}
In this subsection, we consider the $\coh$-graph induced by the incidence graph, and show how its clustering coefficient is affected by the choice of the distributions.
Recall that $x \coh y$ if they belong to a common hyper-edge of any size, so the $\coh$-graph is obtained by replacing every $k$-hyper-edge by a complete graph (a clique) on $k$ vertices.
The clustering coefficient is an important property in the study of social networks, as it measures how much `friends of my friends are friends'.
There are different definitions depending on the context, the one considered here is the local clustering coefficient, $C(v) = M'(v)/ M(v)$, where $M(v)$ is the number of pairs of neighbour vertices of $v$ (vertices at distance $2$ in the incidence graph) and $M'(v)$ is the number of \emph{connected} pairs of neighbour vertices (pairs $v_1$, $v_2$ with $v_1 \coh v_2$).
If $\B(\boldd^v, \boldd^e)$ is a graph ($\boldd^e = \delta_2$), then as observed earlier the number of $\coh$-triangles in the graph is given by Lemma \ref{lem_con} and so is in $O(1)$. 
Therefore in that case the average clustering coefficient goes to $0$, and is $0$ for almost all vertices.
When there are $k$-hyper-edges with $k>2$, the clustering coefficient becomes 
$$C_v = \frac{\sum_{v \in e}{{|e|-1} \choose 2}}{{M \choose 2}} \textup{ with } M=\sum_{v \in e}{|e|-1}.$$
The average clustering coefficient is therefore given by $C = \Ee\boldd^v\Ee{d^e-1 \choose 2}/\Ee{M \choose 2}$ with $d^v \sim \boldd^v$ and $d^e \sim \boldd^e$.
This formula shows that by choosing the correct distributions, it is possible to generate sparse random graphs with a specified clustering coefficient.

\section{Conclusion and open questions}
The logical framework used here allows to extend several classical results on graphs to the very general structure of random hyper-multigraphs,  
solving the problem stated back in \cite{lynch92}: \emph{``Some more interesting problems would be to prove limit laws for structures
with relations of degree greater than 2. Our techniques rely heavily on graph-theoretic concepts, and it is not obvious how to extend them to relations of higher degree''} (edited). 
The solution is simple, and relies on no more than graph-theoretic concepts.

The specific class considered here encompasses a wide range of possible hyper-multigraphs, 
but a wide range of distributions remains to be studied. 
As remarked in \cite{blanchet}, the bipartite configuration model allows to use a larger class of distributions than the simple configuration model, since the maximal degree can be as big as $n-o(n)$, whereas for the simple configuration model the maximal degree must be in $O(\sqrt{n})$ (otherwise the probability that a random configuration is a graph goes to zero).
However for such distributions the methods used here (in particular the local weak convergence) cannot be used. 
When $(**)$ does not hold, as the distributions get heavier tails, subgraphs with positive excess appear, and when $(*)$ does not hold the expected number of cycles becomes unbounded. 
When considering such distributions, there will be non-contiguous sequences defined by distributions with the same supports, and contiguity will also depend on the tail of the distributions.
 
In another direction, only sparse graphs are considered here, but other regimes of interest could be studied, such as sublinear or dense regimes of bipartite graphs. 
Finally, other models of hyper-graphs can be considered in this setting, such as the model considered in \cite{jansonsparsecluster}, or models of multigraphs. 
Some models in the literature may have natural definitions in terms of incidence graphs, while some may be contrived.

\bibliographystyle{plain}
\bibliography{bibfile}

\end{document}